\documentclass[a4paper,12pt]{article}

\usepackage{graphicx}
\usepackage{dcolumn}
\usepackage{bm}

\newcommand{\FPU}{Fermi--Pasta--Ulam\ }
 
\newcommand{\dif}{\,\mathrm {d}}

\renewcommand{\equiv}{\buildrel {\rm def} \over {=} }

\title{ The \FPU system as a model for glasses}
\author{Andrea Carati\thanks{Department of Mathematics, Universit\`a
    degli Studi di  Milano. Via Saldini 50, 20133 Milano -- Italy.
E-mail: \texttt{andrea.carati@unimi.it}} \and
Alberto Maiocchi\footnotemark[1] \and
Luigi Galgani\footnotemark[1] \and \\%
Graziano Amati\thanks{Corso di Laurea in Fisica. Universit\`a degli
  Studi di Milano. Via Celoria 16,  20133 Milano -- Italy.}
}

\date{\today}

\begin{document}
\maketitle
\begin{abstract}
We show that the standard Fermi--Pasta--Ulam system, with a suitable
choice for the interparticle potential, constitutes a model for
glasses, and indeed an extremely simple and manageable one. Indeed, it
allows one to describe the landscape of the minima of the potential
energy and to deal concretely with any one of them, determining the
spectrum of frequencies and the normal modes. A relevant role is
played by the harmonic energy $\mathcal E$ relative to a given
minimum, i.e., the expansion of the Hamiltonian about the minimum up
to second order.  Indeed we find that there exists an energy threshold
in $\mathcal E$ such that below it the harmonic energy $\mathcal E$
appears to be an approximate integral of motion for the whole
observation time. Consequently, the system remains trapped near the
minimum, in what may be called a vitreous or glassy state. Instead,
for larger values of $\mathcal E$ the system rather quickly relaxes to
a final equilibrium state.  Moreover we find that the vitreous states
present peculiar statistical behaviors, still involving the
 harmonic energy $\mathcal E$. Indeed, the vitreous states are described
by a Gibbs distribution with an effective Hamiltonian close to
$\mathcal E$ and with a suitable effective inverse temperature. The
final equilibrium state presents instead statistical properties which
are in very good agreement with the Gibbs distribution relative to the
full Hamiltonian of the system.
\end{abstract}

\section{Introduction}

If one looks at the microscopic equations of motion, 
there is essentially no difference between
a glass and  the corresponding crystal form. They live in the same
phase space and have  the same Hamiltonian, with   a  potential
energy presenting    a huge number of minima. In both  cases the initial
data lie near one of the  potential energy minima, the only peculiarity of
the crystal being that the corresponding minimum is the absolute one.

The glass and the corresponding crystal show 
different  physical behaviors just  because  they
remain trapped for an extremely long time in two different
regions of phase space (two regions about  two different minima,
indeed). A final relaxation to global equilibrium might perhaps occur, but only over
time scales of a huge magnitude, outside the experimental reach. One can
thus conjecture
that  there exist suitable effective integrals of motion, which forbid the system from
exploring the whole a priori available ``energy surface''.

This analogy or some variant of it certainly is at the basis of the
paper \cite{parisi} (see also \cite{parisi2}), in which for the first
time the idea was advanced that an analogy may exist between the
standard \FPU system \cite{FPU} and glasses. Indeed 
it was suggested
that the apparent paradox of lack of energy equipartition observed  in the FPU
system for initial data of
FPU type (long wavelength excitations) might be interpreted as
analogous to the glass--like trapping of a system near a potential
energy minimum. In both cases, a final approach to equilibrium might
occur  over longer time scales. However, apparently such analogy was
no longer pursued.

In the present paper we make explicit such a qualitative analogy,
through 
numerical integrations  of  a FPU system, i.e.,    
a linear chain  of
particles with   nearest--neighbor interaction. The only 
peculiarity of the present model  is that, for the interaction,   a double--well
quartic potential is chosen (Fig. \ref{fig:grafico}). This is 
 easily seen to imply a property which is a characteristic feature of the
 literature on glasses. Namely, that the system   possesses a huge number
 (increasing exponentially 
with the sistem size) of stable
equilibrium points, each corresponding to a disordered configuration
of particles, which are usually said to constitute the 
``potential energy landscape'' of the system.

Obviously, the program of understanding glassy dynamics in terms of
the potential energy landscape is an old one, which goes back at least 
to the work of Goldstein \cite{gold}  and was pursued   more
recently  ib several works   (see for example \cite{altri1}).
The advantage of the  present model seems to be that    in principle it allows one to locate
all the equilibrium points, ascertain their stability, and compute for
each of them the
frequency spectrum  (Fig. \ref{fig:spettro}) together with  the corresponding normal modes of
oscillation (Fig.\ref{fig:modi}). 
So one has an essentially 
complete information on the system, at variance with what occurs with    three-dimensional models.

Our aim is to exploit  such  information in
studying the dynamics of our model  exactly in the spirit of the
standard studies on  the \FPU system.

The numerical simulations show that the system admits
 glassy states: if the initial conditions are chosen sufficiently ``near'' a local
minimum, the trajectory remains trapped in a
region about that point for the duration of the simulations. This
implies  that ergodicity is broken, at least on the times scale of
our simulations. In fact we find that there exists another function,
besides the total Hamiltonian, which remains practically constant
during the evolution. Such a practically constant value  is orders of
magnitudes different from that of  the corresponding   phase average.  
This function is nothing but the sum of the energies of the normal
modes relative to the considered  equilibrium configuration. In the
rest of the paper such a function  will be
called the  ``total harmonic energy'', or simply  the ``harmonic
energy'',  and denoted by $\mathcal{E}$.

It appears that the harmonic energy  $\mathcal{E}$ determines whether  the
system will be trapped or not: if its value is below a certain
threshold  the system does not thermalize, analogously to what occurs in the familiar FPU system.  
Instead,  a transition to a behavior of ergodic type
occurs if the  value of $\mathcal{E}$ is sufficiently raised  
(Figs. \ref{fig:ene512} and \ref{fig:ene4096}).

Obviously, a trajectory might remain trapped about a local potential
minimum because of a trivial reason, i.e.. just in virtue of
conservation of the Hamiltonian.  But this is not what happens in the
case of glasses because, in the thermodynamic limit, there is plenty
of energy available for a particle to leap over a potential barrier.
It is just in order to insure that this happens also in our model that
the double well potential was chosen with a very low barrier between
the two minima (Fig. \ref{fig:grafico}). Thus, in all cases in which
the system appears to be trapped near a vitreous equilibrium point,
the total energy is much larger (by factors of order ten or a hundred,
depending on the system size) than the one needed for a particle to
leap over the barrier.  So, conservation of the Hamiltonian alone
cannot explain the trapping.  Instead, the trapping is apparently
due to the fact that the total harmonic energy $\mathcal{E}$ relative
to the considered minimum is practically a constant of motion in a
neighborhod about it, whereas this no longer occurs for initial data  sufficiently far
away from the stable equilibrium point, i.e., above a certain
threshold in $\mathcal E$.

This is the first result of our paper. A second one pertains to the
distribution of the normal mode energies. Indeed, if computed in the
equilibrium Gibbs state (through a Montecarlo simulation), such
distribution displays a very distinctive character
(Fig.~\ref{fig:modi_equilibrium}). We find that the empirical
distribution observed in the final eauilibrium state above threshldold
agrees very well with the theoretical equilibrium one (see
Fig. \ref{fig:histogibbs512hi}). This is completely at variance with
what occurs in the glassy state (below threshold).  Indeed in such a
case the empirical distribution (see Fig. \ref{fig:histogibbs512low})
follows an exponential law of the type $\exp(-\beta_{eff}E)$, This
suggests that in the glassy state there exists a (quasi--equilibrium)
measure, actually a Gibbs measure $\exp(-\beta_{eff}\tilde H)$, with a
suitable effective Hamiltonian $\tilde H$ close to the total harmonic
energy $\mathcal{E}$ (relative to the considered minimum), and a
suitable effective temperature. This reminds us of what occurs in the
description of the different phases met in phase transitions, as was
particularly stressed by Frenkel \cite{frenkellibro}.

The paper is organized as follows. In Section~\ref{sec:2} we describe
the model, determine an equilibrium point and compute the
corresponding normal modes of oscillation together with the related
spectrum of frequencies.  In Section~\ref{sec:3} we discuss the
statistical features of the equilibrium state.  In Section~\ref{sec:4}
we illustrate the results of our numerical computations, which exhibit
the existence of a glassy state in our model. Some final remarks
are reported in Section~\ref{sec:5}.

\section{The model, the local minima and the corresponding normal modes}\label{sec:2}

\begin{figure}[t]
\center \includegraphics[width=0.75\textwidth]{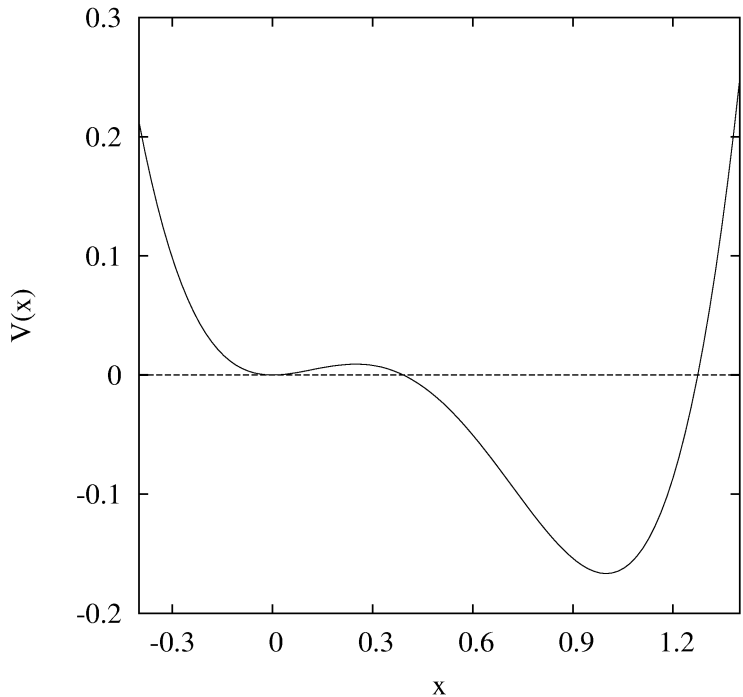}
\caption{\label{fig:grafico} Plot of the two body potential as a
  function of  particle ``distance''.}
\end{figure}
If one considers a linear chain of $N+2$ equal point particles 
with a nearest neighbor interaction and fixed ends, 
 one gets the following Hamiltonian
\begin{equation}\label{eq:ham}
H = \frac 1{2m} \sum_{j=1}^N p_j^2 + \sum_{j=0}^N V(x_{j+1} -x_j)\  ,
\end{equation}
with $x_0 =0$, $x_{N+1}=L$, where $L$ is the total length of the chain. A trivial equilibrium
point is obtained by taking $x_{j+1}-x_j = L/(N+1)$,  which corresponds to
a crystal structure. A more general equilibrium is obtained by
imposing that 
$$
V'(x_{j+1} -x_j) \equiv F(x_{j+1}-x_j) = b \quad \mbox{for all}\quad j \ ,
$$
so that  the two  forces acting on each particle balance. Now, if
$V'(r)$ is monotonic, then the crystal equilibrium is the only
possible one. Otherwise, if   $F^{-1}(b)$ (the inverse image of $b$
under the application of
the function $F$) contains $m>1$ points
$a_1,\ldots, a_m$, then one gets   $m^N$ different equilibrium points,
which are obtained by choosing in all possible ways 
the ``distances'' $ r_j \equiv x_{j+1}-x_j $ between adjacent particles,
taking them  from   the  values $a_l$. As in the
ordered crystal case, $b$ is determined by imposing
$$
\sum r_j = L \ .
$$

Expanding  the potential energy about one of these equilibria up to
second order one gets
\begin{equation}
V_{tot} = \sum V(r_j) + \frac 12 \sum V''(r_j)(q_{j+1} - q_j)^2 + \dots 
\end{equation}
where $q_j$ is the displacement of the  $j$--th particle from its
equilibrium position. This shows that if the values  $r_j$ are such that
$V''(r_j)>0$ for all $j$, then the considered equilibrium point 
is a local minimum, and so is stable. 
We do not discuss here the general case, and in
the present paper we consider only equilibrium points for which the
above condition is satisfied. For example, this certainly occurs if
the two--body potential $V(r)$ presents two minima, i.e., is a double
well potential, and one takes as $r_j$ one of them.
\begin{figure}[t]
\includegraphics[width=0.5\textwidth]{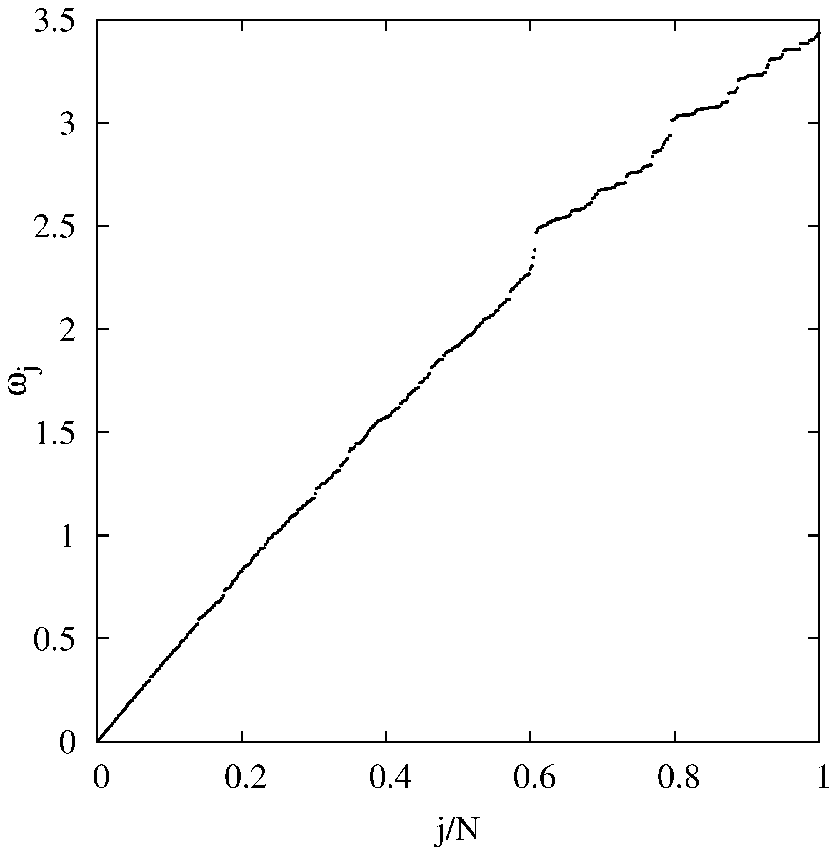}
\includegraphics[width=0.5\textwidth]{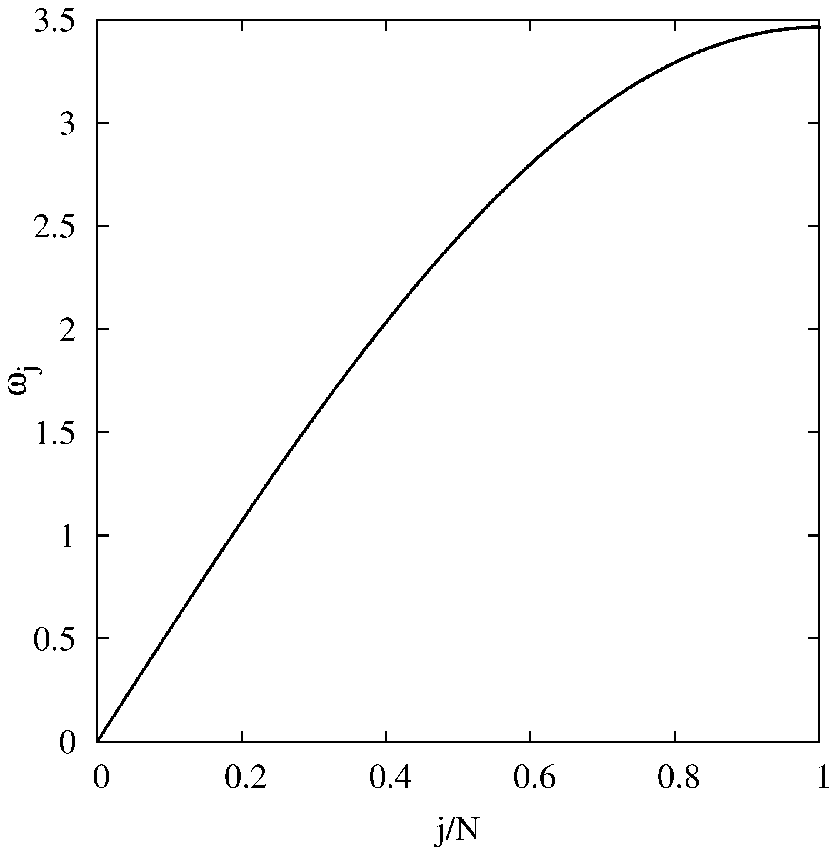}
\caption{\label{fig:spettro} Frequency spectra, i.e., frequencies of
  the normal modes, about a minimum of the potential energy. Left
  panel:  a glass case (local minimum). Right panel: the crystal case
  (absolute minimum). In the glassy case the
  frequencies are reported  in ascending order as a function of the
  index $j/N$.  System size is  $N=512$.}
\end{figure}
\begin{figure}[t]
\includegraphics[width=.5\textwidth]{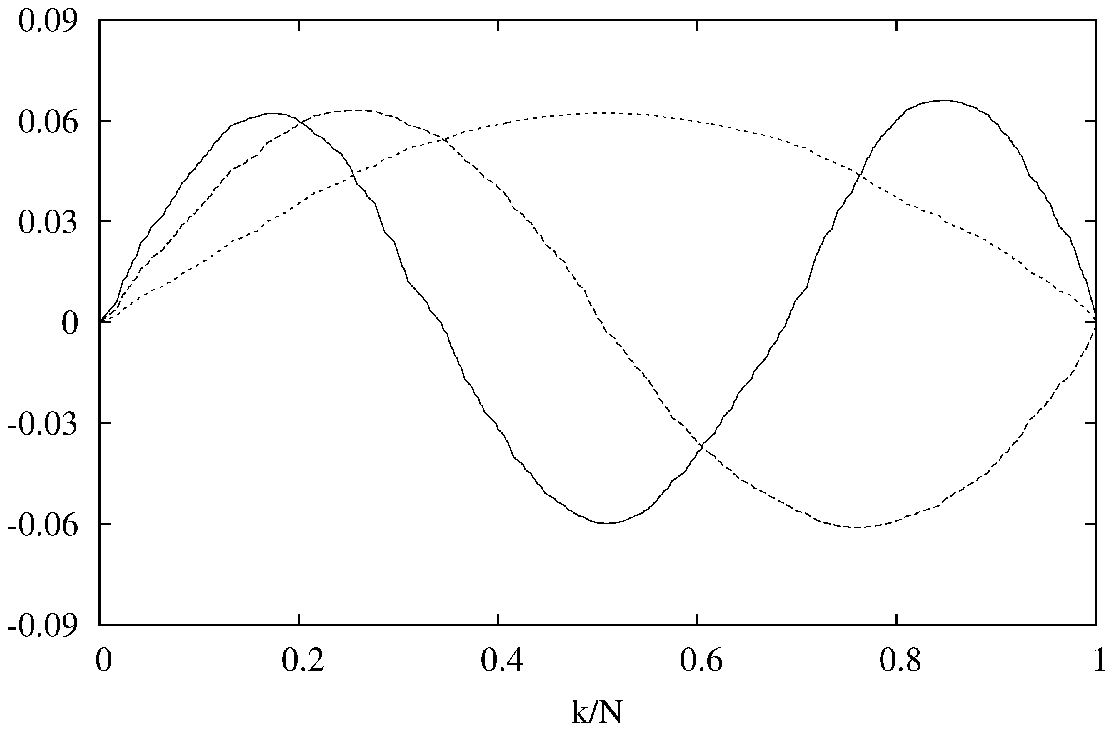}
\includegraphics[width=.5\textwidth]{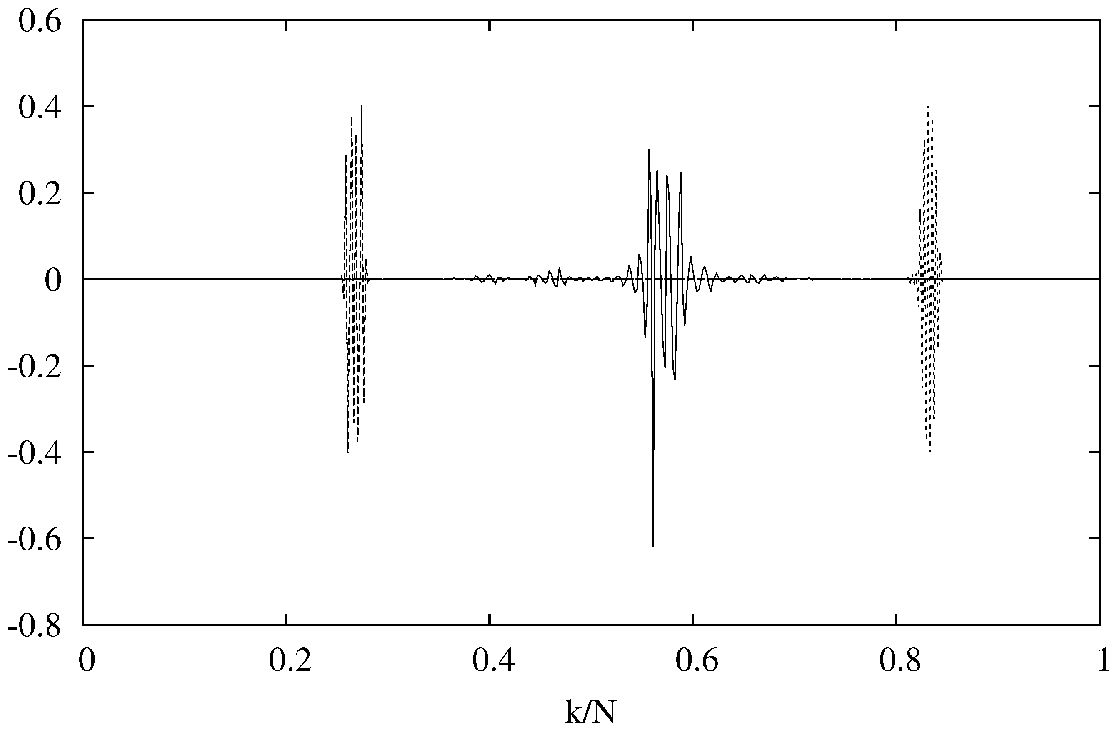}
\caption{\label{fig:modi} Shapes of  the normal modes, namely, of 
  the oscillation amplitudes   of the particles 
 as a function of the particle number normalized to 1, for a disordered
  equilibrium point. System size is $N=512$.  Left panel: the
  three  normal modes of lowest frequencies. The shapes are  virtually the same
  as for the  crystal. 
  Right panel: three normal modes in the higher part of the  frequency
  spectrum (with frequencies $\omega_k=1.93,2.89,3.34$). The
  localization phenomenon is clearly exhibited.
} 
\end{figure}

Clearly each of  such equilibrium points corresponds  to a disordered
structure. Moreover, one    meets here in a natural way with a random frame, because
each of such equilibria can be obtained 
by choosing at random  the distances  $r_j$,  among the
$m$ possible  values $a_l$. 
So one obtains a ``landscape'' of local minima, whose number is
exponentially increasing with the number of particles.

For the purpose of discussing such vitreous states, it is of interest to determine 
the normal modes of a given equilibrium
point. In the disordered case this can be obtained only by numerical
means,  computing the change--of--basis matrix $\hat U$ which
diagonalizes the dynamical matrix. This forces us to make a definite
choice for the two-body potential $V(r)$, which we take  within the
\FPU family $V(r)= {r^2}/2+  \alpha r^3/3+ \beta r^4/4$,
by fixing   $\alpha=-5$ and $\beta=4$, namely,   
\begin{equation}\label{eq:pot} 
V(r)= \frac {r^2}2- \frac  53 r^3+  r^4\ .
\end{equation}
The graph of the potential is displayed in Fig.~\ref{fig:grafico}. As 
one sees, this is an asymmetric double--well potential, with the ratio
 between the heigths of the two barriers equal to
$\approx0.05$, and the distance between the minima equal to 1. With a
potential of such a type, one might guess that a  jump from the higher
minimum to the lower one  easily occurs. However,   our numerical
simulations will show that  this is not the case. 

Having chosen the potential, we determine one among the equilibrium
configurations by choosing at random the distances  between adjacent
particles. We take
$r_j=1$ with probability $0.64$ and $r_j=0$ with probability
$0.36$, in order to insure that the total length be $L$. 
We then compute the change--of--basis matrix $\hat U$ and
determine the normal modes together with  their frequencies. In
Fig.\ref{fig:spettro} we report the spectrum of the disordered system
(left panel).
As the wave number $k$ does not exist in this case, we sort the frequencies
in ascending order, i.e., in such a way that $\omega_{j+1} \ge
\omega_j$. For comparison, we also report in the figure (right panel) the spectrum
of the crystalline system, i.e., of the system for which all $r_j$'s are  equal to
$1$. One sees that the spectra are alike for the first part of the
spectrum, while qualitative differences show up in the second part,
where the disordered spectrum presents some discontinuities.

For what concerns the shapes of the normal modes, some examples are
shown in Fig.~\ref{fig:modi}. The low--frequencies  modes
are delocalized  and quite similar to the
crystalline case (left panel). Instead, the modes  become localized as the frequency
increases, and finally, in the upper part of
the spectrum, they  are localized on only  a very  few sites (right panel). 
This is in agreement with the analytical results reported in \cite{local}.

\section{The statistical approach}\label{sec:3}

We will show in a moment that the vitreous state is not at all
typical, i.e.,  that, if the initial data are taken at random, then the
system is far from any equilibrium point. Here ''random'' means as usual
that the data are extracted according to the Gibbs measure
\begin{equation}
\dif \mu = \frac {e^{-\beta H(p_1,\ldots,x_{N}) }}{Z(\beta)} \dif
p_1\ldots\dif x_N \ .
\end{equation}
Care should be taken in choosing a suitable value of $\beta$, if one  wants that
the  mean energy $U$ falls in a range in which
an exponentially large number of equilibrium points are present. 
With our choice of the
two--body potential, this occurs  if one takes for example
$\beta=6$, which is the value we used in all our numerical
computations. For  such a value of $\beta$ we also computed the average
length of the  chain with free ends, and  fixed the actual
length $L$  to such an average value.  The value $L$, in
turn, determines also the values assigned to the probabilities needed
to build up the glass in the way described  in the previous section.
\begin{figure}[t]
\center \includegraphics[width=\textwidth]{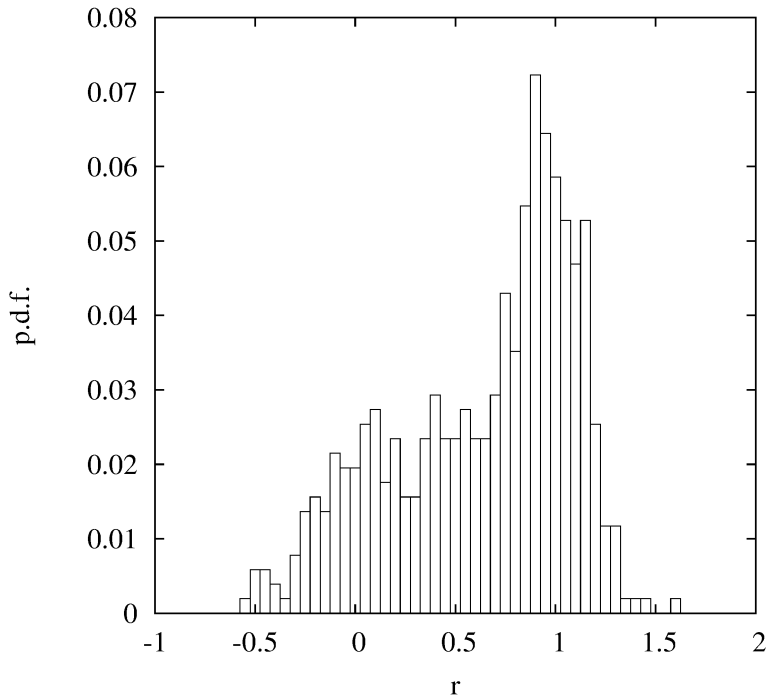}
\caption{\label{fig:posizioni} The probability distribution function
  for the  distance  between adjacent  particles, for a canonical
  distribution with $\beta=6$.}
\end{figure}
\begin{figure}[t]
\center \includegraphics[width=\textwidth]{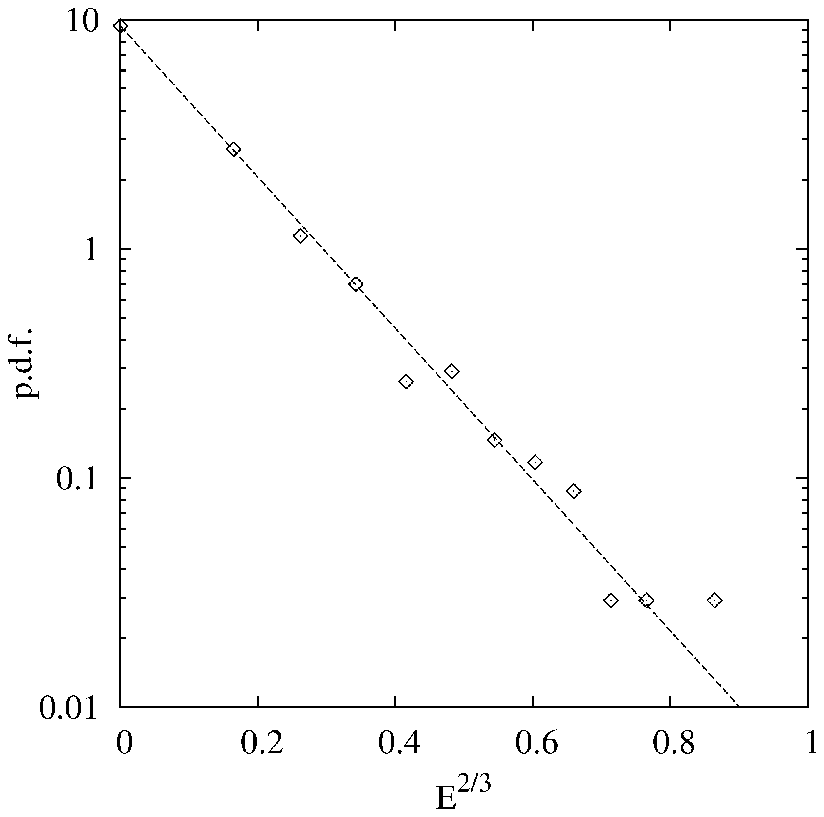}
\caption{\label{fig:modi_equilibrium} For a canonical
  distribution with $\beta=6$, plot of the probability distribution
  function for the energy of the normal modes, versus  $E^{2/3}$, in semilogarithmic scale. The
  continuous line is the function $\exp (-\gamma E^{2/3})$, with $\gamma
  =7.6$. System size is $N=512$. }
\end{figure}

So, having fixed $\beta$, one can extract some data, and construct the
histogram of the  distances $r_j$ 
between  adjacent particles.  The result is shown in
Fig.~\ref{fig:posizioni}.  
As one sees,  the dispersion is very large. In particular, one has  a
maximun at $r=1$ and a relative maximum about $r=0$.
The large dispersion clearly shows that, in a generic configuration,
the particles are not close to any minimum of the potential, so that a vitreous
state in non generic.

A different way of  describing this situation is by looking at 
other quantities.  In particular one can look at the distribution
of the energy of the normal modes relative to a given minimum. This is
shown in Fig.~\ref{fig:modi_equilibrium}.  One finds a
characteristic decay, as  the histogram is well approximated by a
distribution of the type $\exp(-\gamma E^{2/3})$, with
$\gamma=7.6$). This means that the distribution has a heavy tail,
which in turn means that many modes have a big energy, i.e., the point
in phase space
is "far" (in the energy norm) from the considered equilibrium point.
The shape of the histogram is also useful in deciding whether the
system did thermalize or not, as will be better explained below.

Thus a vitreous state can only be constructed by hands in the way
explained in Section~\ref{sec:2}, and not by extracting
it   at random with the Gibbs measure. In other terms, the vitreous
states do not belong to the so--called Boltzmann sea 
(i.e., the set of points of phase space which are ''typical''
with respect to the Gibbs measure). Thus, vitreous states can 
 show up only if the  dynamics is not ergodic, i.e., only if, placing initially the system
near a local  equilibrium point, it later remains ``frozen'' there for a long time 
without entering into the ''Boltzmann sea''. It is in this sense 
that  ergodicity is, as one often 
says, ``broken''.

We suggest two possible ways  to check that the system remains outside the
Boltzmann sea. The first is statistical in nature. One chooses an
initial datum near the equilibrium point representing the glass,
computes the orbit for a certain time, and also computes  the histogram of
the energies of the single normal modes, at the final time. If such histogram  agrees well
with the  one  computed according to  the  Gibbs distribution, in
particular by showing heavy tails, then the system did
thermalize, leaving the neighborhood of the equilibrium
point. Otherwise the system is still frozen in the glassy state.  

A more geometrical approach to control whether the point remains in a
neighborhood of the equilibrium point is the following. Starting again
from an initial datum near the equilibrium point, one computes the
trajectory, and looks at the total harmonic energy $\mathcal{E}$,
i.e., at the sum of the energies of the normal modes, as a function of time.
The set defined by $\mathcal{E}\le \delta$, with $\delta$ much smaller than the
phase average $\langle \mathcal{E} \rangle$, is a (small) ellipsoid
centered at the equilibrium point. If the total harmonic  energy $\mathcal{E}$
approaches its phase average, then the system leaves the neighborhood
and (possibly) does thermalize.

If instead, during the evolution, the total
harmonic energy remains almost constant, close to its initial value,
then the motion is confined in such  an ellipsoid, and so remains close to
the equilibrium point. This shows that the system  remains a ``glass''. 
At the same time  this  shows that
 ergodicity is broken in the standard sense of dynamical system
theory, i.e., there exists a function, independent
of the total energy of the system, whose time average does not
converge to its phase average. 

As said  in the Introduction, the aim of this paper is to  show, through
numerical integrations of  the equation of motion, that this actually
happens, i.e., that there exists a threshold 
 of the  total harmonic  energy $\mathcal E$ such that, if
one starts with a smaller  value of  $\mathcal{E}$  the
system remains  ``frozen'' in the glassy state without thermalizing, up to
the times for which  we can  numerically follow the system. In
this sense  we recover, in the present  setting, the classical results
of  the standard \FPU system.

\section{Numerical results}\label{sec:4}

\begin{figure}
 \center{\includegraphics[width=0.75\textwidth]{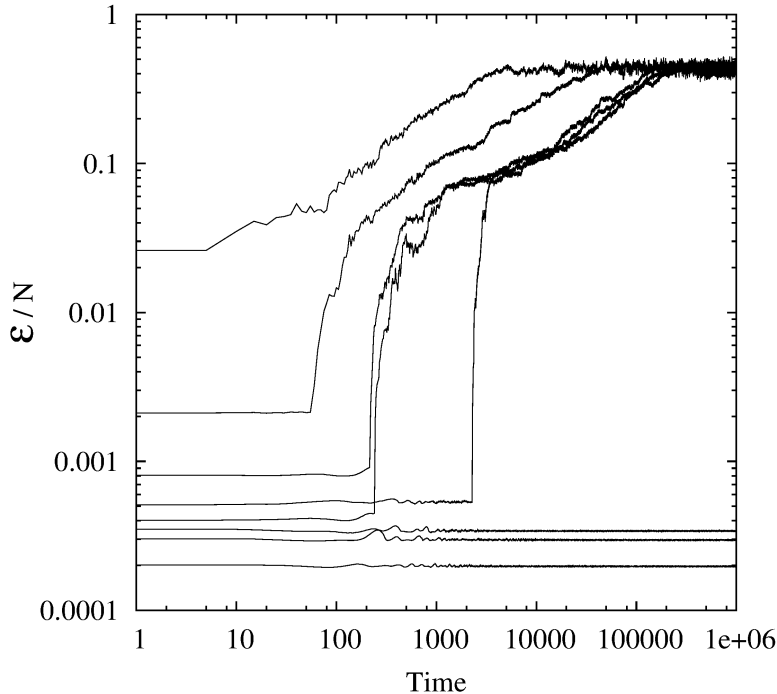} \quad\quad}
\caption{\label{fig:ene512} 
 Total  harmonic energy per particle $\mathcal{E}/N$  as a
 function of time, in logarithmic scale. The curves correspond
  to a system of $N=512$ particles with   initial values of
  $\mathcal{E}/N$ equal to 
  0.0002, 0.0003, 0.00035, 0.0004, 0.0005, 0.0008,  0.002 and 0.006. 
}
\end{figure}
\begin{figure}
\center{\includegraphics[width=0.75\textwidth]{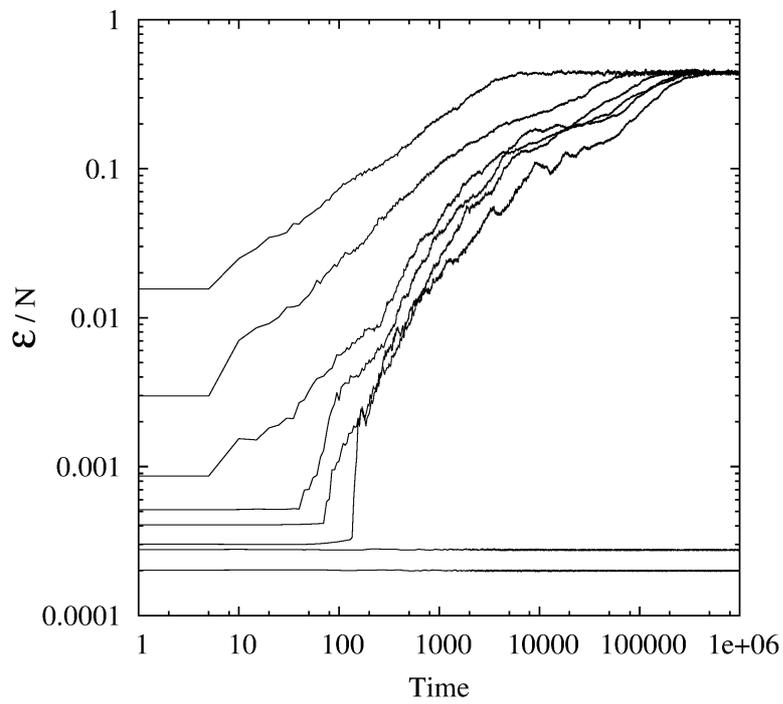} \quad\quad}
\caption{\label{fig:ene4096} 
 Total  harmonic energy per particle $\mathcal{E}/N$  as a
 function of time, in logarithmic scale. The curves correspond
  to a system of $N=4096 $ particles with   initial values of
  $\mathcal{E}/N$ equal to 0.0002, 0.000275,  0.0003,
   0.0004,  0.0005, 0.0008,  0.002 and 0.006. }
\end{figure}
As explained above, we integrated  (by  the standard leap--frog method)
 the equations of motions corresponding to the
Hamiltonian (\ref{eq:ham}), with  a potential $V(r)$ given by
(\ref{eq:pot}). The time step  was chosen equal
to $0.005$,  so as to insure an energy conservation better than
a part over a thousand in all the performed  computations. The total time of each
integration was equal to $10^6$.

We integrated the system for two different numbers $N$ of
particles, namely,  $N=512$ and $N=4096$. 
For each of such  two values of $N$ we constructed a vitreous equilibrium
point as described earlier in Section~\ref{sec:2}. Then the initial
data near it
were  chosen in the familiar way used in the \FPU model, i.e., by 
assigning   energies and phases to the normal modes pertaining to that
equilibrium.

In fact, we chose  to excite only  low--frequency packets of
modes, as done in many numerical studies of the \FPU system (see
\cite{galla} or the recent
work \cite{ponnobenettin}). More precisely, in the case  $N=512$
we excited only the three lowest frequency modes, giving them an
equal share of energy,  while choosing   the phases  at random. In the
case $N=4096$, we excited only the twenty four lowest--frequency 
modes, so that the packet had the same relative width as for 
$N=512$. Again the initial energies were the same for all the excited
modes, with  the phases  chosen at random.
\begin{figure}
\center\includegraphics[width=0.75\textwidth]{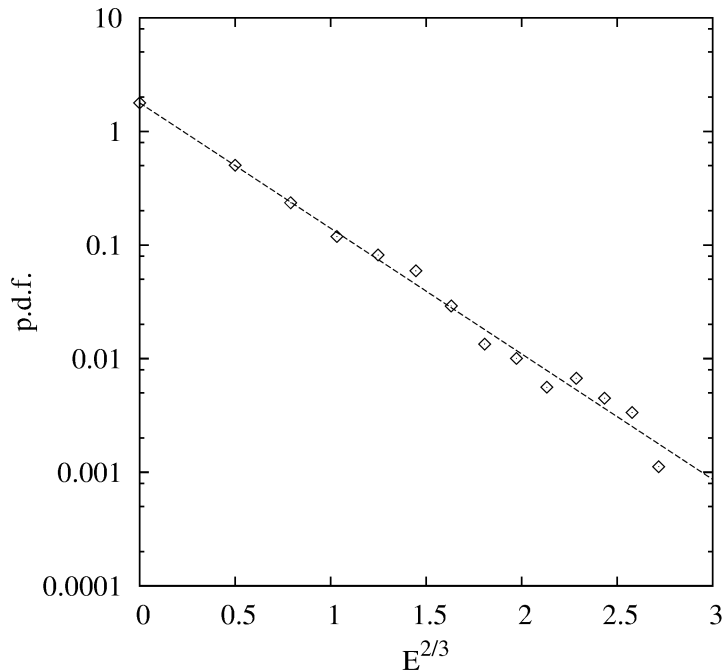}
\caption{\label{fig:histogibbs512hi} Histogram of the distribution
  of the mormal mode energies (diamonds), at the end of
  the numerical integration, in semilogarithmic scale.  
Initial  value of 
  $\mathcal{E}/N$ equal to $0.0008$, a case above  threshold.  The
  values are reported versus $E^{2/3}$, as for the theoretical
  equilibrium p.d.f.  reported in Fig. \ref{fig:modi_equilibrium}. 
  The agreement  is evident.   
   The continuous line  is the graph of the function
   $\exp\, (-\gamma E^{2/3})$, with $\gamma=2.54$. 
}
\end{figure}
\begin{figure}
\center\includegraphics[width=0.75\textwidth]{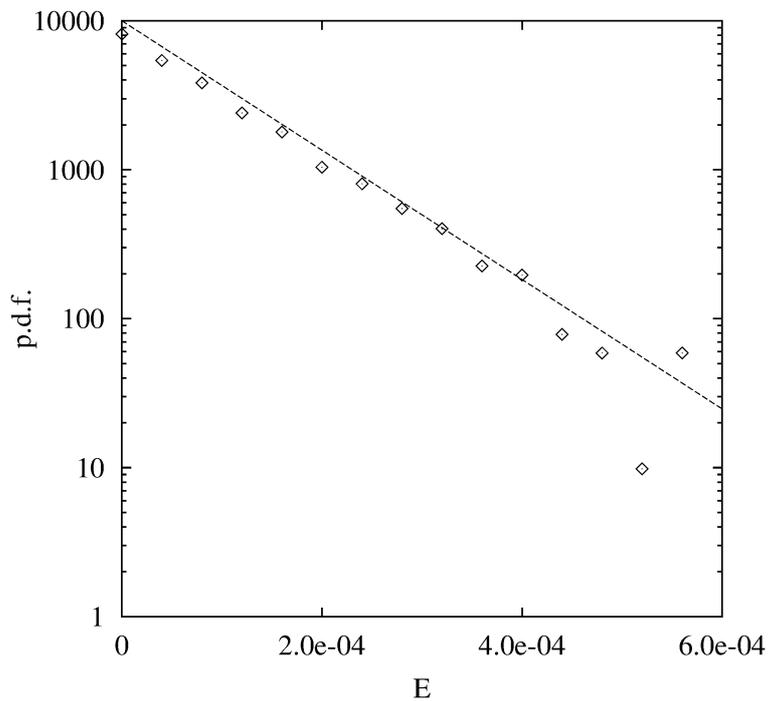}
\caption{\label{fig:histogibbs512low} Histogram of the distribution
  of the mormal mode energies (diamonds), at the end of
  the integration, in semilogarithmic scale.  
Initial value of  
  $\mathcal{E}/N$ is  $0.0001$, a case
  below threshold. Values are reported versus $E$.
  The continuous line 
  is the graph of the
  function $\exp(-\beta_{eff}E)$, with $1/\beta_{eff}=
  \mathcal{E}/N=0.0001$. 
System size is $N=512$. }
\end{figure}

We made different runs in both cases, with the initial total harmonic
energies  chosen in such a way that the  total harmonic energies
per particle
$\mathcal{E}/N$ were essentially the same for the two   values of $N$.  The
results are summarized in the 
Figures~\ref{fig:ene512}-\ref{fig:histogibbs512low}.

In Figure~\ref{fig:ene512}, we report the total harmonic
energy per particle $\mathcal{E}/N$ as a function of time in
logarithmic scale, for the case $N=512$. We report eight runs with
different initial values of $\mathcal{E}/N$ in the range $0.0001 \div
0.006$. One sees that if the initial value of $\mathcal{E}/N$ is above
a threshold (laying between $0.0003\div 0.0004$), the total harmonic
energy $\mathcal{E}$ soon starts increasing, and then reaches, after
some time, an asymptotic value which agrees rather well with the
corresponding phase average computed through the Gibbs measure (with
$\beta =6$). In this case the system did thermalize. Below threshold
things are different: the total harmonic energy $\mathcal{E}$ remains
constant, just fluctuating a bit about its initial value. So below
threshold the system is frozen in a glassy state up to the total
integration time, and ergodicity is broken, at least on such a time
scale. Perhaps the system might thermalize on a longer time scale, but
in any case such time scale has to increase sharply, below the
threshold.

Things are the same also in the case $N=4096$, as one sees in
Figure~\ref{fig:ene4096}. Even in this case the total harmonic energy
$\mathcal{E}$ goes asymptotically to its phase average if the initial
value of $\mathcal{E}/N$ is above a threshold (laying now between
$0.00025 \div 0.0003$) while remaining essentially constant below such
a threshold.  Notice that the threshold of $\mathcal{E}/N$ appears to
depend on the number $N$ of particles, even if not in a quite strong
way. However, as the threshold could also exhibit a dependence on the
chosen local minimum, the dependence of the threshold on the number
$N$ of particles needs to be more carefully investigated. We leave
this task for future studies, and in this paper we content ourselves
with indicating that the values of the threshold (expressed in terms
of energy per particle) have roughly the same order of magnitude in
the two cases.

A very interesting fact shows up if one investigates the distribution
of the energies of the single normal modes. This is shown in
Figure~\ref{fig:histogibbs512hi}--\ref{fig:histogibbs512low}, which
give the histograms for the values of the energies at the final time
of integration, The figure refers to $N=512$ (the case $N=4096$ gives
the same results, and so the histograms are not reported here).  The
first figure refers to an initial value of $\mathcal{E}/N$ equal to
$0.0008$, a case above threshold. Here the system did thermalize, and
in fact one sees that the histogram decays as a stretched exponential,
with the same power $E^{2/3}$ as the one computed at equilibrium. 
Things, instead, are completely different
below threshold. This is shown in Figure \ref{fig:histogibbs512low},
which corresponds to an initial value of $\mathcal{E}/N$ equal to
$0.0001$. Now one sees that the distribution is of the type
$\exp(-\beta_{eff}E)$, i.e., of Gibbs type. However, one has now an
effective (quadratic) Hamiltonian instead of the true Hamiltonian,
and moreover an effective inverse temperature $\beta_{eff}$ with
$1/\beta_{eff}=\mathcal{E}/N$, as might have been expected.  So it
appears as if the glassy state could be described by a measure of
Gibbs type with an effective Hamiltonian and an effective
temperature. Such a measure is different from the Gibbs one relative
to the true hamiltonian, which instead describes the final equilibrium
state very far from the glassy state.
 
\section{Final comments}\label{sec:5}

We have shown that, in our \FPU model  with a double--well interparticle
potential, it is possible to build a
vitreous state which is stable for a  very long time. Thus the
system fails to exhibit an ergodic behavior
on large time scales, in  very close analogy
with what  observed in the original FPU paper.  

Now, perhaps this might have been forecast on the basis of the present
theoretical understanding of the FPU model, particularly (see for
example \cite{BCM}) for what concerns the slowing down of relaxation
to equilibrium. However, the same can not be said concerning the other
main result found here for the glassy states, which came as a
surprise. Namely, the Gibbs--like form of the histogram of the
distribution of the mode energies. This seems to indicate that the
measure which describes the glassy state has  first of all to be of
Gibbs type, with however both an effective Hamiltonian and an
effective inverse temperature $\beta_{eff}$, in place of the true
Hamiltonian and the true $\beta$. Actually, a phenomelogical approach
of this type was taken in the physical literature (see for example
\cite{quarzosilice}).  However, a clear theoretical
understanding of this point is apparently lacking at the moment.

Another interesting point is that, below threshold,  the total harmonic energy
$\mathcal{E}$ appears to be a  conserved quantity, 
independest of  the total Hamiltonian $H$. Now,  in the standard studies on the FPU model, i.e., 
those concerned with the crystal state, it is usually assumed 
that $\mathcal{E}$ is conserved \emph{because}  of  conservation of
the Hamiltonian $H$, the two  quantities being  very close to each
other  in that case. 
So the present results cast some doubts on that belief.
The interesting point is that, if $\mathcal{E}$ corresponds to a
conserved quantity independent of $H$, then  the  criterion  usually
employed for checking thermalization in the FPU system  
(namely,  the
occurring or not of equipartition
of the normal mode energies)  should be reconsidered. Indeed in the
case of glasses such a criterion would lead to estimates of the
relaxation times  much  lower than those found here.




\end{document}